\documentclass[aps,prl,reprint,superscriptaddress,amsmath,amssymb]{revtex4-1}
\usepackage{graphicx}
\usepackage{dcolumn}
\usepackage{bm}

\begin{document}

\title{Unambiguous separation of the inverse spin Hall and anomalous Nernst Effects within a ferromagnetic metal using the spin Seebeck effect}

\author{Stephen M. Wu}
	\email{swu@anl.gov}
\affiliation{%
Materials Science Division, Argonne National Laboratory, Argonne, Illinois 60439, USA
}%

\author{Jason Hoffman}
\affiliation{%
Materials Science Division, Argonne National Laboratory, Argonne, Illinois 60439, USA
}%

\author{John E. Pearson}
\affiliation{%
Materials Science Division, Argonne National Laboratory, Argonne, Illinois 60439, USA
}%

\author{Anand Bhattacharya}%

\affiliation{%
Materials Science Division, Argonne National Laboratory, Argonne, Illinois 60439, USA
}%

\date{\today}

\begin{abstract}
The longitudinal spin Seebeck effect is measured on the ferromagnetic insulator Fe$_3$O$_4$ with the ferromagnetic metal Co$_{0.2}$Fe$_{0.6}$B$_{0.2}$ (CoFeB) as the spin detector. By using a non-magnetic spacer material between the two materials (Ti), it is possible to decouple the two ferromagnetic materials and directly observe pure spin flow from Fe$_3$O$_4$ into CoFeB. It is shown, that in a single ferromagnetic metal the inverse spin Hall effect (ISHE) and anomalous Nernst effect (ANE) can occur simultaneously with opposite polarity. Using this and the large difference in the coercive fields between the two magnets, it is possible to unambiguously separate the contributions of the spin Seebeck effect from the ANE and observe the degree to which each effect contributes to the total response. These experiments show conclusively that the ISHE and ANE in CoFeB are separate phenomena with different origins and can coexist in the same material with opposite response to a thermal gradient. 
\end{abstract}

\maketitle

	Within the last decade, a large amount of attention has focused on the generation and detection of pure spin currents without associated charge currents. This is done through the use of newly discovered phenomena such as, the spin Hall effect (SHE)  \cite{d1971possibility,hirsch1999spin,kato2004observation}, spin pumping \cite{tserkovnyak2002spin}, and the spin Seebeck effect (SSE) \cite{uchida2010spin,uchida2010observation,uchida2012thermal,weiler2012local}. The longitudinal SSE is particularly attractive, because of the large spin currents that can be generated through only the application of heat \cite{weiler2013experimental}. This effect can be described as a spin current  generated in the direction parallel to a thermal gradient applied across an insulating ferromagnet without any flow of charge. When this ferromagnet is coupled to another material, pure spin current can flow into the adjacent material and be detected through the inverse spin Hall effect (ISHE). Since the discovery of the ISHE in metallic ferromagnetic systems \cite{miao2013inverse}, interest has been generated in using ferromagnetic metals as spin current detectors instead of the typical high-Z paramagnets. Due to the detector layer being a ferromagnet itself, additional steps must be taken to separate the contributions from the ISHE and the anomalous Nernst effect (ANE). The ANE presents itself as a voltage proportional to magnetization, generated perpendicular to both magnetization and thermal gradient within a ferromagnetic metal. The SSE detected by the ISHE is phenomenologically identical except the voltage is now proportional to magnetization of the ferromagnetic source layer. 
	
	The initial work on this subject by Miao et al. has shown that the ISHE exists in the ferromagnetic metal permalloy in direct contact with the ferromagnetic insulator yttrium iron garnet (YIG) \cite{miao2013inverse}. This group attributes the additional voltage measured without a spin current blocking layer to the ISHE. But, since the two ferromagnetic materials are in direct contact, there is a direct exchange interaction between the two systems, and thus the magnetic properties of the metallic ferromagnetic layer are modified. Due to this fact, it becomes less straightforward to separate the ISHE in permalloy from a modified ANE due to exchange interactions at the interface. 
	
	Additionally, the inseparability of the ISHE and the ANE has been a long-standing source of problems in SSE experiments due to proximity magnetic interactions between spin current sources and spin current detectors\cite{huang2012transport,kikkawa2013longitudinal,lu2013pt,
geprags2012investigation,qu2013intrinsic}. In some spin detector materials such as Pt, which are close to a magnetic instability, some portion of the detector material will become magnetic and contribute a voltage due to the ANE. Within these systems it is not possible to clearly separate out the contribution from each effect without resorting to separate control experiments due to their almost identical response to thermal gradient. 
	
\begin{figure}[b]
\includegraphics[width=3.4in,trim =1in 1.5in 1.5in 2.75in,clip=true]{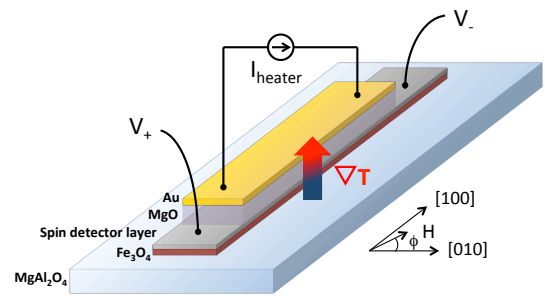}
\caption{\label{fig1} A schematic of a typical device structure with an on-chip heater separated by an electrically insulating MgO layer barrier. Magnetic field is applied at an angle $\phi$ to measure the SSE when a thermal gradient is applied through the heater. }
\end{figure}	 
	 
\begin{figure*}[t]
\includegraphics[width=7in,trim =0in 0in 0in 0in,clip=true]{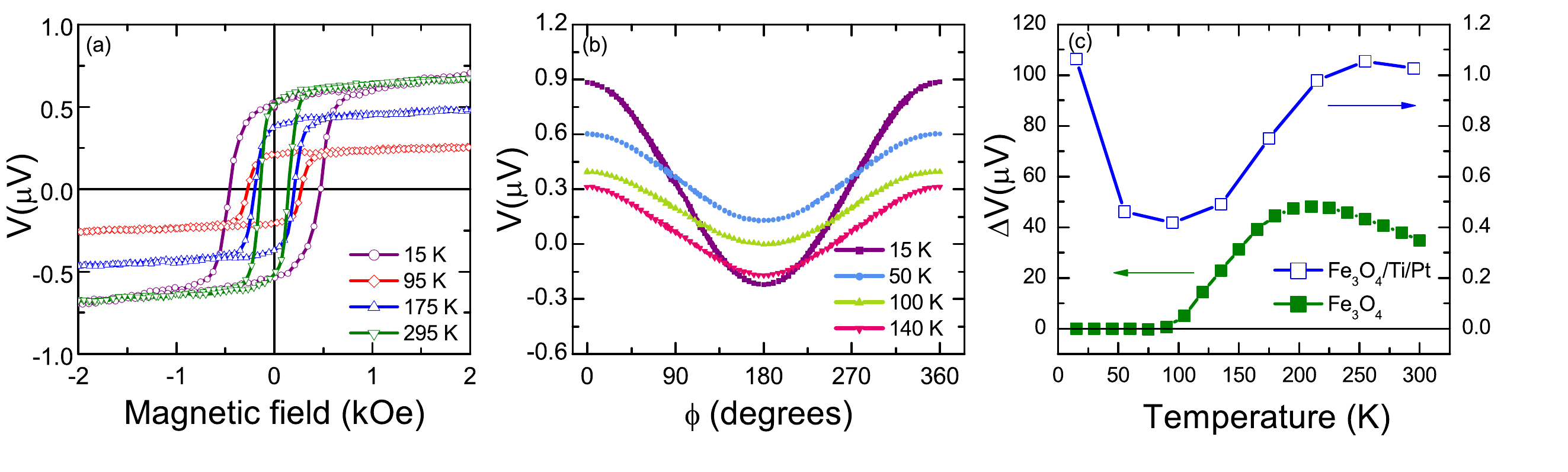}
\caption{\label{fig2} Measurements performed on a Fe$_3$O$_4$(30 nm)/Ti(1.3 nm)/Pt(5 nm) structure on a MgO (100) substrate. Voltage vs. magnetic field are measured in (a) for $\phi=0^\circ$. A magnetic field independent offset is removed for clarity,likely arising from the charge Seebeck effect due to device asymmetry. Angular dependent measurements of the voltage are performed with a 4 V$_{pp}$ voltage bias at a static 5000 Oe, as shown in (b). Temperature dependent magnitude of V vs. H, as defined as the voltage difference between saturation magnetization states, at $\phi=0^\circ$ are presented in (c) for a 5 V$_{pp}$ voltage bias in both devices with and without Pt. }
\end{figure*}	

	 In this Letter, we provide a solution to both these issues through the use of Co$_{0.2}$Fe$_{0.6}$B$_{0.2}$ (CoFeB), a ferromagnetic metal where the ANE and ISHE occur simultaneously with opposite polarity, thus allowing for unambiguous differentiation between the two effects. In our experiment, CoFeB is deposited on ferromagnetic Fe$_3$O$_4$ with a non-magnetic Ti spacer layer in between. Fe$_3$O$_4$ is an insulator below a well-known charge ordering phase transition at $\sim$120 K referred to as the Verwey transition. Ti was chosen for its small voltage response in SSE experiments as measured by other groups\cite{ishida2013observation}, as well as our own. Fe$_3$O$_4$ was chosen because its coercivity is much larger than CoFeB. These facts allow for the clear decoupling of exchange interactions between the two magnetic materials, while still allowing for spin current to flow.  
	 
	 Using ozone assisted molecular beam epitaxy, 60 nm of Fe$_3$O$_4$ was grown on MgAl$_2$O$_4$ (100) or MgO (100) substrates. Details of the growth and characterization of these films are outlined in earlier work \cite{liu2013non}. Following oxide growth, using photolithography and liquid nitrogen cooled Ar$^+$ ion milling, a 10 $\mu$m x 800 $\mu$m strip of Fe$_3$O$_4$ was patterned by etching the entire film to the substrate.  Ti (15 nm)/CoFeB (3.5 nm)/Ti (5 nm) was then deposited by sputtering at room temperature. The top Ti layer is used both as a capping layer and to eliminate any Rashba effects due to inversion asymmetry from the interfaces. On top of this entire structure, MgO (100 nm)/Au (100 nm) was deposited using e-beam evaporation to serve as a direct on-chip heater separated by a thick insulating dielectric. A schematic of the device is shown in Fig. \ref{fig1}. By using an on-chip heater, the experiment is contained within a single micropatterned device that allows for a more direct probe of magnetism at the device level, as well as easy integration into conventional cryostat systems.  To perform measurements, the chip is loaded into a Quantum Design Physical Property Measurement System cryostat, and a sinusoidal current is applied to the heater layer using a function generator. To eliminate any parasitic effects and increase measurement sensitivity, we take advantage of the fact that the voltage generated due to the SSE and the ANE are proportional to the amount of heat applied due to Joule heating, such that $V\propto I_{heater}^2$. Thus, if the heater current is:
\begin{equation}\label{heater1}
I_{heater}\propto sin(\omega t),
\end{equation}
then the subsequent measured voltage is:
\begin{equation}\label{heater2}
V\propto \frac{1}{2}(1 -cos(2\omega t)).
\end{equation}
By ignoring the constant term and using a lock-in amplifier to detect the 90$^{\circ}$  out-of-phase component at the $2\omega$ frequency, we are able to detect voltages due to heating down to the 5 nV level. Experiments were typically carried out at 99 Hz. 

	As a control, a similar device was fabricated with the Ti/CoFeB/Ti stack replaced by Ti (1.3 nm)/Pt (5 nm) stack in a 5 $\mu$m x 400 $\mu$m device. Above the Verwey transition, there are contributions from both the ANE in Fe$_3$O$_4$ and the ISHE in Ti/Pt. Both effects cause a voltage to develop when a thermal gradient is applied, since $\vec{E}_{ISHE}\propto \vec{J_S} \times \hat{\sigma}$ and $\vec{E}_{ANE}\propto \nabla T \times \vec{M}$. $\vec{E}_{ISHE}$ and $\vec{E}_{ANE}$ are the electric fields produced by both  effects, $\vec{J_S}$ is the spin current, and $\vec{M}$ is the magnetization of the ferromagnet. Because the spin current $\vec{J_S}$ is directly related to both thermal gradient $\nabla T$, and magnetization $\vec{M}$, both effects present a similar voltage response. 
	
	 To distinguish the two effects, we perform a series of device characterization measurements. Using a function generator, a 5 V$_{pp}$ signal was applied across a 50 ohm load resistor and a 50 ohm top heater in series, for an approximate power applied across the heater at room temperature of $\sim$15.6 mW$_{rms}$. Temperature dependent results are presented in Fig. \ref{fig2}a. These V vs. H curves match well with the expected magnetization hysteresis curves for typical Fe$_3$O$_4$ grown on MgO (100). Angular dependent measurements taken with a static magnetic field of 5000 Oe are presented in Fig. \ref{fig2}b. This is enough to fully saturate the magnetization of the underlying film at all temperatures and thus the voltage response follows a standard cos($\phi$) dependence resulting from the cross product term in equations describing $\vec{E}_{ISHE}$ and $\vec{E}_{ANE}$. Finally, to separate out the contributions from the ANE and SSE, the magnitude of each V vs. H response at $\phi=0^\circ$ is presented as a function of temperature for both Fe$_3$O$_4$/Ti/Pt and a separate device only containing Fe$_3$O$_4$ without any spin detector layer. From Fig. \ref{fig2}c, it is possible to see that in the Fe$_3$O$_4$ only device, the ANE signal disappears below the Verwey transition, likely due to carrier freeze-out\cite{ramos2013observation}. In the Fe$_3$O$_4$/Ti/Pt device, there is a recovery in signal at low temperatures due to the SSE. We have excluded the possibility of proximity magnetic interactions in Pt by inserting the extra layer of Ti. We can gather from this evidence that at low temperatures we eliminate the contribution from the Fe$_3$O$_4$ ANE and can separately examine the effect of the additional metallic layers only. The large difference in magnitude of the two responses in Fig. \ref{fig2}c can be attributed to the large resistivity difference between Fe$_3$O$_4$ and Ti/Pt. Using an analysis similar to the analysis in the first demonstration of the SSE in Fe$_3$O$_4$ \cite{ramos2013observation}, we find that our ANE signal is reduced to ~1\% when Ti/Pt is added.
	 
	 In this experiment there is not a direct probe for the thermal gradient across the thin film. Since all the measurements were performed under a constant heater voltage bias, while the heater resistance changes with temperature, the heater power also varies with temperature ($\sim$40\%). Concurrently, the thermal conductivities of the substrate, Fe$_3$O$_4$, and the platinum film are also changing with temperature. Despite these issues, measurements match up remarkably well with previously published work by Ramos et al. on the spin Seebeck effect in Fe$_3$O$_4$/Pt structures \cite{ramos2013observation}. This suggests that our method produces qualitatively similar results to those measured by applying a constant temperature gradient across the film and substrate, where the same problem of concurrently changing thermal conductivities also occurs \footnote{The one notable difference is that in the work by Ramos et al. the voltage response in the device with Pt is larger than the device without. This may be related proximity magnetic effects due to the omission of the Ti spacer layer.}.
	
\begin{figure}[h]
\includegraphics[width=3.4in,trim =0.75in 1in 1in 1in,clip=true]{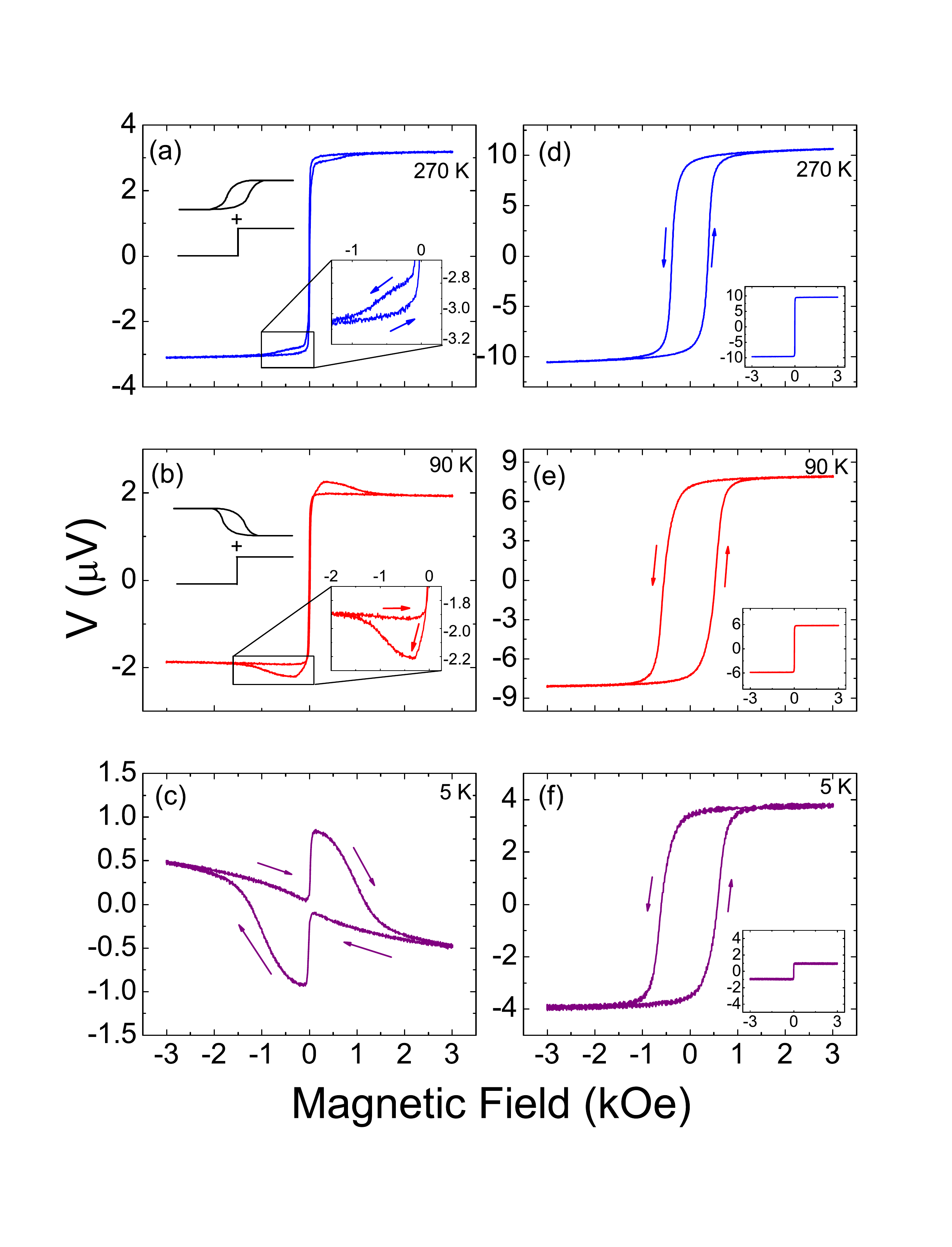}
\caption{\label{fig3} Voltage vs. magnetic field measurements ($\phi=0^\circ$) for multiple devices. The power applied to each device at 270 K, 90K, and 5 K are 3.6 mW$_{rms}$, 2.7 mW$_{rms}$, and 2.2 mW$_{rms}$ respectively. For (a)-(c), the device is Fe$_3$O$_4$ (60 nm)/Ti (15 nm)/CoFeB (3.5 nm)/Ti (5 nm). For (d)-(f), the device is Fe$_3$O$_4$ (60 nm)/CoFeB (3.5 nm)/Ti (5 nm). For (d)-(f) inset, the device is CoFeB (3.5 nm)/Ti (5 nm). All devices were grown on MgAl$_2$O$_4$ (100) substrates. }
\end{figure}	 
	
	Using the same technique, now we replace the Ti/Pt film with a Ti/CoFeB/Ti trilayer. By choosing the first Ti layer to be 15 nm, the Fe$_3$O$_4$ and CoFeB become magnetically decoupled, and the individual responses from each can be separated. In Fig. \ref{fig3}a-c the V vs. H curves at $\phi=0^\circ$ are shown at different temperatures under a constant 2.8 V$_{pp}$ voltage bias across the heater and a 50 ohm load resistor. In these different temperature regimes, it is possible to resolve effects from Fe$_3$O$_4$ and CoFeB separately, because the two have largely different coercive fields. In typical samples on MgAl$_2$O$_4$, H$_c$ for CoFeB is $\sim$10 Oe, while H$_c$ for Fe$_3$O$_4$ is $\sim$500 Oe at room temperature. The coercivity is larger for Fe$_3$O$_4$ grown on MgAl$_2$O$_4$ than for samples grown on MgO, presented in Fig. \ref{fig2}, due to the larger lattice mismatch.
	
	At 270 K, there are contributions from the ANE in CoFeB, the ANE in Fe$_3$O$_4$, and the ISHE in CoFeB due to the SSE from Fe$_3$O$_4$ (Fig. \ref{fig3}a). In this case, the ANE in Fe$_3$O$_4$ is larger than the ISHE in CoFeB, and the overall voltage response shows a positive correlation to the magnetization for both the small and large coercivity magnets. Below the Verwey transition, the contribution from the ANE in Fe$_3$O$_4$ is eliminated and we now only measure the voltage response in CoFeB. At 90 K, the ISHE effect in CoFeB is clear and has a negative correlation to the magnetization in Fe$_3$O$_4$, opposite to the ANE signal for CoFeB which remains positive with respect to the magnetization of CoFeB (Fig. \ref{fig3}b). At 5 K, since the temperature-dependent magnitudes of the ANE and SSE change relative to each other, the ISHE becomes larger than the ANE signal in CoFeB (Fig. \ref{fig3}c). A separate control experiment with 15 nm of Ti on Fe$_3$O$_4$ has shown that the voltage response due to the Ti layer is less than 10 nV, close to the sensitivity limit of our measurement setup.
	
	When the same experiment is performed on a sample where CoFeB is directly coupled to the Fe$_3$O$_4$, as in the work by Miao et al. with YIG/Permalloy \cite{miao2013inverse}, the results are very different (Fig. \ref{fig3}d-f). Here, instead of being able to individually resolve the different contributions, exchange interactions at the interface create a strongly coupled single magnetic system with properties from both individual magnets. Thus, the V vs. H curve is directly proportional to the magnetization of this combined magnet, and there is no opposite polarity SSE response that can be resolved. The insets in Fig. \ref{fig3}d-f show the V vs. H curves for a CoFeB/Ti bilayer without Fe$_3$O$_4$, and the measured ANE signal shows that the coercivity of this material remains as small as in the decoupled system in Fig. \ref{fig3}a-c. 
	
\begin{figure}[h]
\includegraphics[width=3.4in,trim =0.75in 0in 1in 0.75in,clip=true]{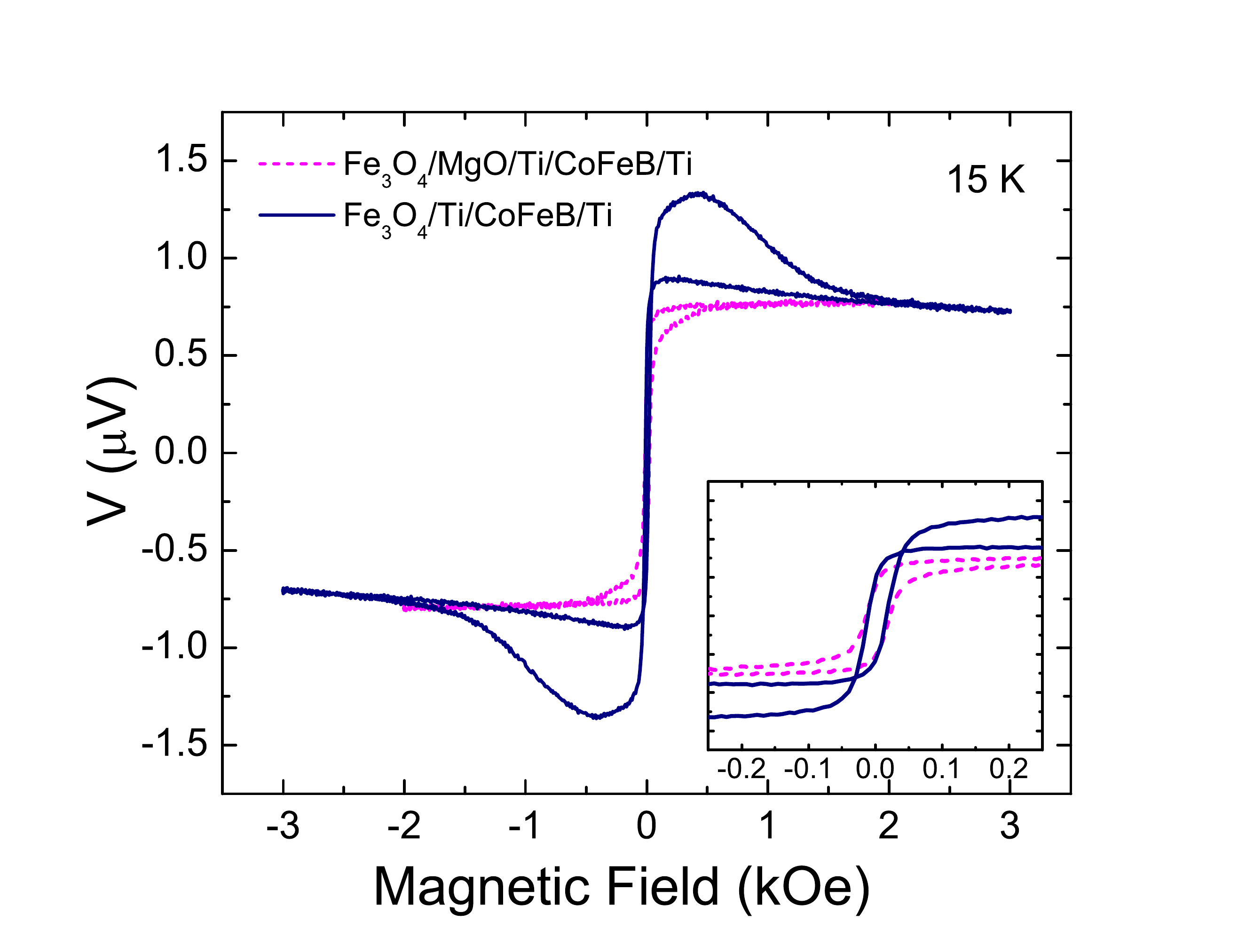}
\caption{\label{fig4} Two voltage vs. magnetic field ($\phi=0^\circ$) measurements for devices. With a spin current blocking MgO layer: Fe$_3$O$_4$ (60 nm)/MgO (10 nm)/Ti (5 nm)/CoFeB (3.5 nm)/ Ti (5 nm). Without an MgO layer: Fe$_3$O$_4$ (60 nm)/Ti (15 nm)/CoFeB (3.5 nm)/ Ti (5 nm). Both devices grown on MgAl$_2$O$_4$ (100) substrates. The applied power in the measurement with the MgO layer and without the MgO layer is 1.2 mW$_{rms}$ and 2.2 mW$_{rms}$, respectively.  }
\end{figure}

	Finally, to further illustrate the importance of spin current in this system, the V vs. H curve at $\phi=0^\circ$ is measured for a system with a 10 nm MgO blocking layer between the Fe$_3$O$_4$ and the Ti/CoFeB/Ti stack (Fig. \ref{fig4}). When compared to a sample without the MgO blocking layer, it is clear that the extra response is due to spin current flow into CoFeB from the underlying Fe$_3$O$_4$. Here the coercivity of the CoFeB layer is the same, as seen in the inset, suggesting that there is no change in its magnetic properties with and without the MgO layer.

	By looking at the physical origin of both the ANE and the ISHE, it is possible to see that they do not need to have the same polarity. The anomalous Hall effect (AHE) in ferromagnetic systems is due to the preferential transverse scattering of spin-polarized charge carriers, creating a charge current perpendicular to both magnetization and longitudinal current. In the absence of spin-polarized carriers, the same mechanisms cause a spin current to arise for the SHE \cite{hoffmanspinhallmetals}. Since the ISHE is the reciprocal effect of the SHE, the AHE and the ISHE must also be connected through this relation \cite{kimura2007room}. However, the polarity of the Nernst effect does not need to be the same as either the AHE or the ISHE, since the effect depends on the {\em derivative} of the Hall conductivity with respect to energy at the Fermi level following the Mott relation \cite{wang2001onset,lee2004anomalous}:
\begin{equation}\label{mott}
\alpha_{yx}=\frac{\pi^2k_b^2T}{3e}(\frac{\partial \sigma_{yx}}{\partial E})_{E_F},
\end{equation}
where $\alpha$ is defined through the relation:
\begin{equation}\label{mott2}
\vec{J}=\sigma\vec{E}+\alpha(-\nabla T).
\end{equation}
The Mott relation is equally valid for both the ordinary and the anomalous components of the Nernst effect \cite{pu2008mott,miyasato2007crossover}.

	We have shown unambiguously that the ISHE can coexist with the ANE in a single ferromagnetic material with opposite responses to thermal gradient. Since previous experiments relied on directly coupled ferromagnetic materials \cite{miao2013inverse}, which was shown in our experiment to have a clear effect on the magnetization of both magnets, our result is more definitive. Additionally, if there were proximity  magnetic interactions coupling CoFeB to Fe$_3$O$_4$, the result could be read out as a contribution to the measured voltage with a positive response to Fe$_3$O$_4$ magnetization. In our experiment, the SSE causes an ISHE signal with a large negative response to Fe$_3$O$_4$ magnetization, larger than the CoFeB ANE at low T, thus ruling out that this additional effect could be due to such proximity effects. 
	
	We have shown in this Letter that using an on-chip local heater it is possible to examine the longitudinal SSE on a microscopic scale, which is equivalent to measuring the SSE using a static thermal gradient on a larger sample. Using this method, it is possible to separate the ANE and the ISHE in a single ferromagnetic metal (CoFeB) by flowing pure spin current through a non-magnetic spacer material (Ti) from a high coercivity insulating ferromagnet (Fe$_3$O$_4$). This unambiguously proves that the two effects can coexist within the same material and do not share a mutual origin. We are able to clearly examine the degree to which each effect contributes to the total response without resorting to separate control experiments, since our result produces two qualitatively different responses within the same measurement. Since the origin of the SHE shares a relation to the AHE, by further characterizing the relation between the AHE and the ISHE, it may be possible to use the AHE as a guide to find new ferromagnetic materials that exhibit large spin to charge conversion, leading to new novel materials for use in thermal spintronics applications. 

\begin{acknowledgments}
All authors acknowledges support of the U.S. Department of Energy (DOE), Office of Science, Basic Energy Sciences (BES), Materials Sciences and Engineering Division. The use of facilities at the Center for Nanoscale Materials, was supported by the U.S. DOE, BES under contract No. DE-AC02-06CH11357. The authors also thank Axel Hoffmann for valuable discussion and insight. 
\end{acknowledgments}

\end{document}